\documentclass{article}
\usepackage{dcase2022_techrep,amsmath,graphicx,url,times,booktabs, tabularx}
\usepackage{xcolor}
\usepackage{amssymb,amsfonts}
\usepackage{array}

\title{A Robust framework for sound event localization and detection on real recordings}

\name{Jin Sob Kim$^\ast$, Hyun Joon Park$^\ast$, Wooseok Shin$^\ast$, Sung Won Han$^\ast$$^\ast$ \thanks{$^\ast$Equal contribution.}\thanks{$^\ast$$^\ast$Corresponding author.}}
\address{School of Industrial and Management Engineering, Korea University, Seoul, Republic of Korea}
\begin{document}

\ninept
\maketitle

\begin{sloppy}

\begin{abstract}
This technical report describes the systems submitted to the DCASE2022 challenge task 3: sound event localization and detection (SELD). The task aims to detect occurrences of sound events and specify their class, furthermore estimate their position. Our system utilizes a ResNet-based model under a proposed robust framework for SELD. To guarantee the generalized performance on the real-world sound scenes, we design the total framework with augmentation techniques, a pipeline of mixing datasets from real-world sound scenes and emulations, and test time augmentation. Augmentation techniques and exploitation of external sound sources enable training diverse samples and keeping the opportunity to train the real-world context enough by maintaining the number of the real recording samples in the batch. In addition, we design a test time augmentation and a clustering-based model ensemble method to aggregate confident predictions. Experimental results show that the model under a proposed framework outperforms the baseline methods and achieves competitive performance in real-world sound recordings.
\end{abstract}

\begin{keywords}
DCASE2022, Sound event localization and detection, Framework, Test time augmentation, Ensemble
\end{keywords}

\section{Introduction}
\label{sec:intro}
Sound event localization and detection (SELD) aims identification of both the sound event occurance (SED) and the direction of arrival from the sound source (DOA). Since the localization and detection joining task challenge launched in DCASE2019 \cite{Adavanne2018_JSTSP, politis2020overview}, the conceptual formulation of the SELD task has been established in DCASE2021 based on activity-coupled Cartesian direction of arrival (ACCDOA) \cite{Shimada2021}. Unlike the task has been trained and evaluated on the emulated sound scenes in control up to previous iterations, DCASE2022 engages the real sound scape recordings along with strong temporal and spatial annotations bringing the task to the real-world problem \cite{Politis2022starss22}. The challenge also inspires the exploitation of external data to solve the lack of real-world labeled data.

In this report, we introduce the total training framework achieving the generality of model performance on the real sound recordings through utilizing the emulations from the external data sources. Emulated sound data consist of the synthesis of the individual audio samples corresponding to the task-predefined classes, extracted from the external datasets \cite{45857, fonseca2022FSD50K, politis2020dataset, politis2021dataset, piczak2015dataset, juan_j_bosch_2014_1290750, nagatomo2022wearable}, along with spatial room impulse responses (SRIR) and spatial ambient noises (SNoise) \cite{politis_archontis_2022_6408611}. Considering the difference between the size of real sound recordings dataset and of emulated sounds, the framework exploits the down-sized bootstrapping of emulations per every training epoch. The framework also employees the data augmentation techniques to train further generalized model.

Additionally, aiming at the increment of robustness in model prediction, the clustering-based test time rotation-augmentation ensemble method of ACCDOA is designed in this work. The method allows the effective aggregation of the predictions from confident majorities whilst selective rejection of the abnormal candidates.

\section{Proposed Method}
\label{sec:method}

In this section, we describe a proposed framework for SELD, and it consists of feature extraction, data augmentation, external mix, network, and test time augmentation. The overall process of the framework is shown in Figure \ref{fig:framework}.

\begin{figure*}[!t]
\centering
\includegraphics[width=\textwidth]{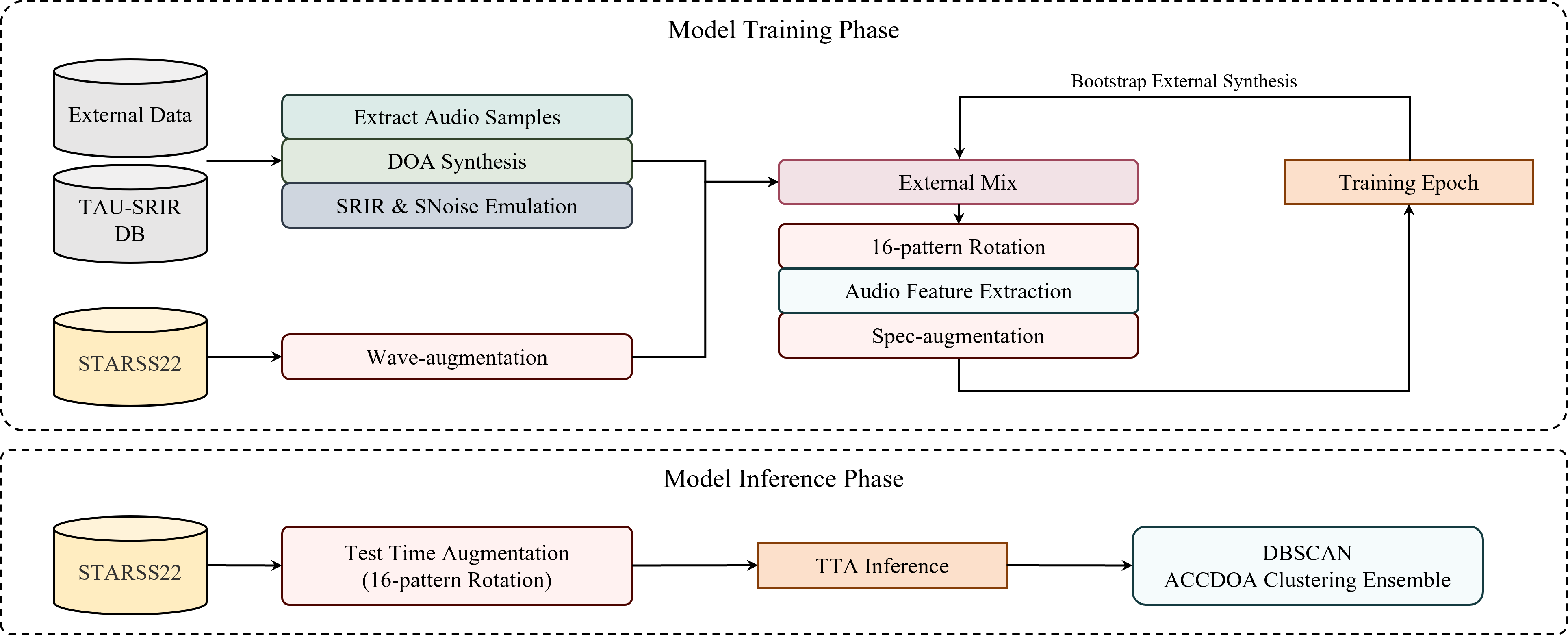}
\caption{Proposed framework overview.}
\label{fig:framework}
\end{figure*}

\subsection{Features}
\label{sec:feature}
We use first-order Ambisonic (FOA) format signals, and multichannel log mel-spectrograms and FOA intensity vectors are used as frame-wise features. The parameters $(sr,nfft,hop,window)=(24000,2048,600,1200)$ are set for Short-time Fourier transformation.

\subsection{Data augmentation}
\label{sec:aug}

We adopt three types of augmentation including each wave and spectrogram augmentation for guaranteeing the generalized performance on the test set. Regarding wave augmentation on the audio inputs, we adjust the gain, shift the pitch, and apply the band-pass filter on the audio. Furthermore, we rotate the directions of events in a sound source as in \cite{mazzon2019first}. The rotation function sets are pre-defined as 16 patterns to change azimuth and elevation angles of a sound source. The rotation function sets are the combinations of channel swapping and channel sign inversion on the $X$, $Y$, and $Z$ axis of FOA inputs. Finally, we use spectrogram augmentation \cite{park2019specaugment}, which maskes time and frequency information after log mel-spectrogram transformation on the audio inputs.

\subsection{External mix}
\label{sec:datamix}

To keep the model fit in real-world scenario contexts while taking advantage of various audio samples from external datasets, dataset mixing technique is adopted to consist model training dataset. The technique balances the size of each dataset on the model training phase, between the small real recording set and the large emulated scenarios. The balancing of two different datasets is conducted as follows: the half of the model training dataset consists of the real-world sound scenarios, where it is maintained to the end of the training phase. The other half comprises with the scenarios down sampled from the emulated sounds, and the sampling occurs dynamically at the start of the every training epoch. 

\subsection{Network architecture}
\label{sec:network}

Figure \ref{fig:model_archtiecture} presents the process of our SELD architecture consisting of an encoder, a self-attention pooling, and a decoder used in this study.
First, we adopt the squeeze-and-excitation residual networks \cite{hu2018squeeze} (SE-ResNet), which have recently been applied to audio classification \cite{yang2018se,shim2020audio}, as the SELD encoder. Our SE-ResNet-34 is the same original SE-ResNet-34, except we use average pooling instead of the stride convolution in the second and third blocks.
In addition, to preserve the frame-wise localization information of the input audio, we aggregate frequency dimensions into channel dimensions using the variant of self-attention pooling \cite{cai2018exploring,chung2020defence}. For the SELD decoder, we employ two bidirectional GRU layers followed by layer normalization and tangent hyperbolic activation. Then, the SELD output is obtained by applying two fully-connected layers followed by tangent hyperbolic activation.

\begin{figure}[t]
  \centering
  \centerline{\includegraphics[width=\columnwidth]{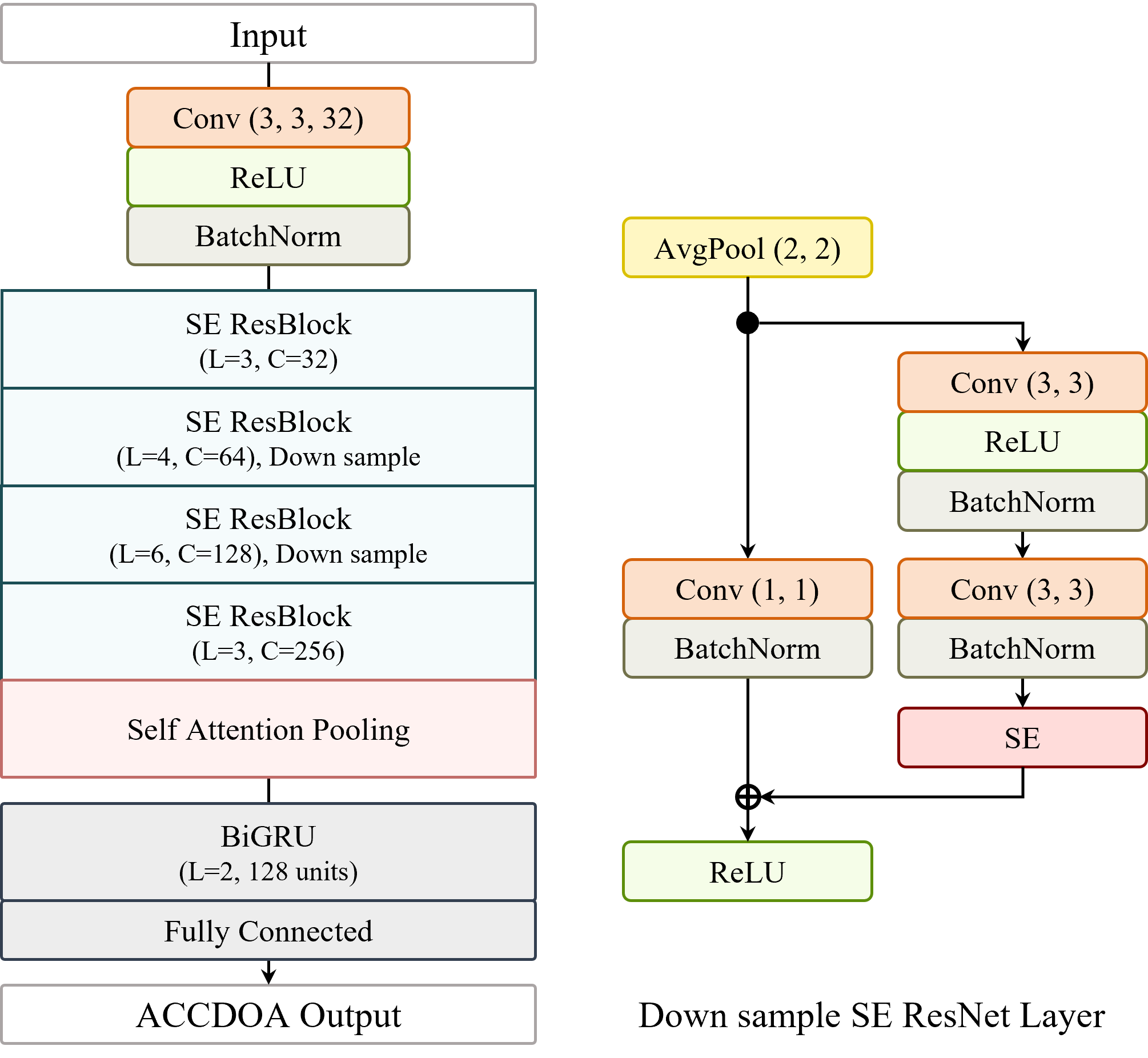}}
  \caption{Model architecture.}
  \label{fig:model_archtiecture}
\end{figure}

\subsection{Test time augmentation}
\label{sec:tta}
To maximize the generalization ability of models in the real recording test phase, we adopt a test time augmentation (TTA). TTA is widely used in computer vision to increase the robustness and performance of models \cite{cohen2019certified, krizhevsky2012imagenet, shanmugam2021better}. Although they proposed better augmentation policies and aggregation methods, they are not appropriate to apply on SELD due to the different predicted outputs. In the specific, the unknown number of events and the presence of coordinates information make challenging to apply TTA on SELD.

To utilize TTA on SELD, we propose a clustering-based aggregation method to obtain confident predicted outputs and aggregate them. We take 16 pattern rotation augmentation for test time augmentation, making 16 predicted outputs, that is candidates. To obtain confident aggregated outputs, we use DBSCAN \cite{ester1996density} for clustering candidates, only when the candidates have the same class and the number of them is over the threshold. The epsilon of DBSCAN is the distance calculated by the threshold of unification for inference in degrees. By exploiting DBSCAN, the outliers of the candidates can be excluded and it is possible to divide the events which have the same class but are differently located. After acquiring clusters of candidates, we take an average to aggregate them. When the number of the aggregated outputs is over three tracks, we select top-3 outputs depending on the weight which is calculated by multiplication between the number of candidates per cluster and their norm. Furthermore, we fit the model on rotated validation set to maximize TTA effect. Finally, we apply this strategy to cross-validation models for the submission of the evaluation dataset. 

\section{EXPERIMENTS}
\subsection{Dataset} \label{subsec:dataset}
The proposed method is evaluated on STARSS22 \cite{Politis2022starss22}. To synthesize emulated sound scenarios from the external data, we use class-wise audio samples extracted from six external datasets, which are AudioSet \cite{45857}, FSD50K \cite{fonseca2022FSD50K}, DCASE2020 and 2021 SELD datasets \cite{politis2020dataset, politis2021dataset}, ESC-50 \cite{piczak2015dataset}, IRMAS \cite{juan_j_bosch_2014_1290750}, and Wearable SELD \cite{nagatomo2022wearable}. $W$ axis of the ambisonic audio format is extracted for the external data that originally consisted in multichannel recording format aiming the SELD task. As the same way in former SELD task challenges, extracted audio samples are synthesized through SRIR and SNoise from TAU-SRIR DB \cite{politis_archontis_2022_6408611} emulating the spatial sound environment.

Before synthesizing the external data, we split the external audio samples from each dataset into four folds for the cross validation. Then, we synthesized two sets for each fold, one keeps the original class distribution of external samples and the other balances the distribution via down sampling the larger class population.

At the developmental stage, the model is trained on the real recording training set and the last external emulation fold while it is validated and also tested on the dev-test set of the given STARSS22 dataset. On the other hand, two types of cross validating divisions are additionally set for the evaluation stage, the first is based on class event-wise stratified split and the other is based on the room-wise split. Both of the splits consist to have four folds each, and the corresponding external emulation folds are considered as external data to be fed the model in training framework.

\subsection{Experimental setup}
We used the official metrics \cite{politis2020overview} to evaluate our SELD system. Similar to previous DCASE SELD challenges, DCASE 2022 SELD challenge includes four evaluation metrics: location-dependent error rate ($ER_{20^\circ}$), location-dependent F-score ($F_{20^\circ}$),  class-dependent localization error ($LE_{CD}$), and localization recall metric ($LR_{CD}$). In contrast to the previous challenges, macro-average is used as a class averaging method for $ER_{20^\circ}$.

During the training, we splitted the audio inputs into 5-second segments with 4-second overlapped. An AdamW optimizer with a learning rate of $1\times 10^{-3}$ was employed. We calibrated the learning rate in half if the validation score did not reduce after 25 epochs. We set a batch size to 128 and total epochs to 200.

\begin{table}[ht]
\centering
\caption{Experimental results of our systems for the dev-test set.}
\label{table:main}
\resizebox{.473\textwidth}{!}{%
\begin{tabular}{lccccc}
\toprule
Ver. & \# Params. & ER$_{20^{\circ}}$ & F$_{20^{\circ}}$ & LE$_{CD}$ & LR$_{CD}$ \\
\midrule
Baseline              & 0.60M & 0.71 & 21\% & 29.30$^{\circ}$ & 46\%  \\   
- w/o Ext. Data       &   -   & 0.84 & 16\% & 43.15$^{\circ}$ & 31\%  \\   
+ w/ Larger Ext. Data &   -   & 0.67 & 35\% & 18.86$^{\circ}$ & 45\%  \\   
\midrule
SE-ResNet+GRU         & 6.04M & 0.59&  40\% & 17.52$^{\circ}$ & 50\% \\   
- w/o Ext. Data       &   -   & 0.83 & 17\% & 36.86$^{\circ}$ & 30\% \\   
+ w/ Larger Ext. Data &   -   & 0.64 & 44\% & 17.53$^{\circ}$ & 65\% \\   
\midrule
SE-ResNet+GRU  & 6.04M & 0.59&  40\% & 17.52$^{\circ}$ & 50\% \\   
+ Augmentation &   -   & 0.55&  46\% & 15.78$^{\circ}$ & 54\% \\   
+ External Mix &   -   & 0.45&  54\% & 14.53$^{\circ}$ & 69\% \\   
+ TTA          &   -   & 0.43&  58\% & 12.86$^{\circ}$ & 69\% \\   
\bottomrule
\end{tabular}%
}
\end{table}



\subsection{Experimental results}
\label{sec:experiments}

In Table \ref{table:main}, we conducted an ablation study to validate the influence of the proposed framework. The results of the first (Baseline model) and second (SE-ResNet) blocks show the effect of the external dataset on performance. The first row of each block, which uses dev-train data and FSD50K (Baseline synthesized data), presented that SE-ResNet significantly outperforms the baseline model; however, the second row of each block using dev-train only showed poor performance for both models. In addition, both models showed additional performance gains when using the seven external datasets mentioned in \ref{subsec:dataset}. In the last block, we found that significant improvements were obtained from three components (Augmentation, External Mix, and TTA). Among them, the external mix method contributed more to the performance improvement than the other methods.

\subsection{Submission}
\label{sec:submit}

Table \ref{table:submission} shows the setups of our submitted systems. Submission \#1 and \#2 are the system trained on development training set with validated and tested solely via dev-test set. The first system is validated without considering the TTA while the second system is validated for all 16-patterns of rotating augmentation. System \#3 and \#4 are trained and validated from each cross validation division aforementioned in Section \ref{subsec:dataset}, class-wise stratified split and room-wise split respectively. TTA was considered for validating models from both systems.

\begin{table}[ht]
\centering
\caption{Experimental results of our submitted systems.}
\label{table:submission}
\resizebox{.473\textwidth}{!}{%
\begin{tabular}{lll}
\toprule
System & \# Params. & Description \\
\midrule
Submission \#1 & 6.04M & SE-ResNet+GRU (Aug \& Ext. mix) \\
Submission \#2 & 6.04M & SE-ResNet+GRU (Aug \& Ext. mix \& TTA)  \\
\midrule
Submission \#3 & 24.16M & SE-ResNet+GRU (CV \& Aug \& Ext. mix \& TTA) \\
Submission \#4 & 24.16M & SE-ResNet+GRU (CV \& Aug \& Ext. mix \& TTA) \\
\bottomrule
\end{tabular}%
}
\end{table}

\section{Conclusion}
\label{sec:conclusion}

In this report, we presented our method for DCASE2022 task 3: sound event localization and detection. We proposed a framework for SELD which composes of data augmentation, external mix, and test time augmentation. The augmentations including the transformation of wave feature, directional rotation, and spectrogram augmentation were effective to increase the diversity of samples. The external mixing technique, which balances the size of the dataset between real-world and synthetic sounds during the training phase, contributed to the improvement of performance on the real-world recordings. Regarding test time augmentation, we designed a clustering-based test time augmentation, which is appropriate for SELD task, and it showed a performance gain. Experimental results demonstrated that the proposed robust framework for SELD with our SE-ResNet-34 can guarantee the generalized performance on the real-world recordings even when the train set of real-world recordings was not enough.

\section{ACKNOWLEDGMENT}
\label{sec:ack}
This research was supported by Brain Korea 21 FOUR. This research was also supported by a Korea TechnoComplex Foundation Grant (R2112651, R2112652).

\bibliographystyle{IEEEtran}
\bibliography{paper}

@misc{Politis2022starss22,
    author = "Politis, Archontis and Shimada, Kazuki and Sudarsanam, Parthasaarathy and Adavanne, Sharath and Krause, Daniel and Koyama, Yuichiro and Takahashi, Naoya and Takahashi, Shusuke and Mitsufuji, Yuki and Virtanen, Tuomas",
    title = "STARSS22: A dataset of spatial recordings of real scenes with spatiotemporal annotations of sound events",
    year = "2022",
    eprint = "2206.01948",
    archiveprefix = "arXiv",
    primaryclass = "eess.AS",
    url = "https://arxiv.org/abs/2206.01948"
}

@inproceedings{45857,
title	= {Audio Set: An ontology and human-labeled dataset for audio events},
author	= {Jort F. Gemmeke and Daniel P. W. Ellis and Dylan Freedman and Aren Jansen and Wade Lawrence and R. Channing Moore and Manoj Plakal and Marvin Ritter},
year	= {2017},
booktitle	= {Proc. IEEE ICASSP 2017},
address	= {New Orleans, LA}
}

@article{fonseca2022FSD50K,
  title={{FSD50K}: an open dataset of human-labeled sound events},
  author={Fonseca, Eduardo and Favory, Xavier and Pons, Jordi and Font, Frederic and Serra, Xavier},
  journal={IEEE/ACM Transactions on Audio, Speech, and Language Processing},
  volume={30},
  pages={829--852},
  year={2022},
  publisher={IEEE}
}

@inproceedings{politis2020dataset,
    author = "Politis, Archontis and Adavanne, Sharath and Virtanen, Tuomas",
    title = "A Dataset of Reverberant Spatial Sound Scenes with Moving Sources for Sound Event Localization and Detection",
    booktitle = "Proceedings of the Detection and Classification of Acoustic Scenes and Events 2020 Workshop (DCASE2020)",
    address = "Tokyo, Japan",
    month = "November",
    year = "2020",
    pages = "165--169",
    url = "https://dcase.community/workshop2020/proceedings"
}

@inproceedings{politis2021dataset,
    author = "Politis, Archontis and Adavanne, Sharath and Krause, Daniel and Deleforge, Antoine and Srivastava, Prerak and Virtanen, Tuomas",
    title = "A Dataset of Dynamic Reverberant Sound Scenes with Directional Interferers for Sound Event Localization and Detection",
    booktitle = "Proceedings of the 6th Detection and Classification of Acoustic Scenes and Events 2021 Workshop (DCASE2021)",
    address = "Barcelona, Spain",
    month = "November",
    year = "2021",
    pages = "125--129",
    isbn = "978-84-09-36072-7",
    doi. = "10.5281/zenodo.5770113",
    url = "https://dcase.community/workshop2021/proceedings"
}

@inproceedings{piczak2015dataset,
  title = {{ESC}: {Dataset} for {Environmental Sound Classification}},
  author = {Piczak, Karol J.},
  booktitle = {Proceedings of the 23rd {Annual ACM Conference} on {Multimedia}},
  date = {2015-10-13},
  url = {http://dl.acm.org/citation.cfm?doid=2733373.2806390},
  doi = {10.1145/2733373.2806390},
  location = {{Brisbane, Australia}},
  isbn = {978-1-4503-3459-4},
  publisher = {{ACM Press}},
  pages = {1015--1018}
}

@dataset{juan_j_bosch_2014_1290750,
  author       = {Juan J. Bosch and
                  Ferdinand Fuhrmann and
                  Perfecto Herrera},
  title        = {{IRMAS: a dataset for instrument recognition in 
                   musical audio signals}},
  month        = sep,
  year         = 2014,
  publisher    = {Zenodo},
  version      = {1.0},
  doi          = {10.5281/zenodo.1290750},
  url          = {https://doi.org/10.5281/zenodo.1290750}
}

@inproceedings{nagatomo2022wearable,
  title={Wearable SELD dataset: Dataset for sound event localization and detection using wearable devices around head},
  author={Nagatomo, Kento and Yasuda, Masahiro and Yatabe, Kohei and Saito, Shoichiro and Oikawa, Yasuhiro},
  booktitle={ICASSP 2022-2022 IEEE International Conference on Acoustics, Speech and Signal Processing (ICASSP)},
  pages={156--160},
  year={2022},
  organization={IEEE}
}

@dataset{politis_archontis_2022_6408611,
  author       = {Politis, Archontis and
                  Adavanne, Sharath and
                  Virtanen, Tuomas},
  title        = {{TAU Spatial Room Impulse Response Database (TAU- 
                   SRIR DB)}},
  month        = apr,
  year         = 2022,
  publisher    = {Zenodo},
  doi          = {10.5281/zenodo.6408611},
  url          = {https://doi.org/10.5281/zenodo.6408611}
}

@article{Adavanne2018_JSTSP,
    author = "Adavanne, Sharath and Politis, Archontis and Nikunen, Joonas and Virtanen, Tuomas",
    journal = "IEEE Journal of Selected Topics in Signal Processing",
    title = "Sound Event Localization and Detection of Overlapping Sources Using Convolutional Recurrent Neural Networks",
    year = "2018",
    volume = "13",
    number = "1",
    pages = "34-48",
    doi = "10.1109/JSTSP.2018.2885636",
    issn = "1932-4553",
    month = "March",
    url = "https://ieeexplore.ieee.org/abstract/document/8567942"
}

@article{politis2020overview,
    author = "Politis, Archontis and Mesaros, Annamaria and Adavanne, Sharath and Heittola, Toni and Virtanen, Tuomas",
    title = "Overview and Evaluation of Sound Event Localization and Detection in DCASE 2019",
    journal = "IEEE/ACM Transactions on Audio, Speech, and Language Processing",
    volume = "29",
    pages = "684--698",
    year = "2020",
    publisher = "IEEE",
    url = "https://ieeexplore.ieee.org/abstract/document/9306885"
}

@inproceedings{Shimada2021,
    Author = "Shimada, Kazuki and Koyama, Yuichiro and Takahashi, Naoya and Takahashi, Shusuke and Mitsufuji, Yuki",
    title = "ACCDOA: Activity-Coupled Cartesian Direction of Arrival Representation for Sound Event Localization and Detection",
    month = "June",
    year = "2021",
    address = "Toronto, Ontario, Canada",
    booktitle = "IEEE International Conference on Acoustics, Speech and Signal Processing (ICASSP)"
}

@inproceedings{cohen2019certified,
  title={Certified adversarial robustness via randomized smoothing},
  author={Cohen, Jeremy and Rosenfeld, Elan and Kolter, Zico},
  booktitle={International Conference on Machine Learning},
  pages={1310--1320},
  year={2019},
  organization={PMLR}
}

@article{krizhevsky2012imagenet,
  title={Imagenet classification with deep convolutional neural networks},
  author={Krizhevsky, Alex and Sutskever, Ilya and Hinton, Geoffrey E},
  journal={Advances in neural information processing systems},
  volume={25},
  year={2012}
}

@inproceedings{shanmugam2021better,
  title={Better aggregation in test-time augmentation},
  author={Shanmugam, Divya and Blalock, Davis and Balakrishnan, Guha and Guttag, John},
  booktitle={Proceedings of the IEEE/CVF International Conference on Computer Vision},
  pages={1214--1223},
  year={2021}
}

@article{mazzon2019first,
  title={First order ambisonics domain spatial augmentation for DNN-based direction of arrival estimation},
  author={Mazzon, Luca and Koizumi, Yuma and Yasuda, Masahiro and Harada, Noboru},
  journal={arXiv preprint arXiv:1910.04388},
  year={2019}
}

@article{park2019specaugment,
  title={Specaugment: A simple data augmentation method for automatic speech recognition},
  author={Park, Daniel S and Chan, William and Zhang, Yu and Chiu, Chung-Cheng and Zoph, Barret and Cubuk, Ekin D and Le, Quoc V},
  journal={arXiv preprint arXiv:1904.08779},
  year={2019}
}

@inproceedings{hu2018squeeze,
  title={Squeeze-and-excitation networks},
  author={Hu, Jie and Shen, Li and Sun, Gang},
  booktitle={Proceedings of the IEEE conference on computer vision and pattern recognition},
  pages={7132--7141},
  year={2018}
}

@inproceedings{yang2018se,
  title={Se-resnet with gan-based data augmentation applied to acoustic scene classification},
  author={Yang, Jeong Hyeon and Kim, Nam Kyun and Kim, Hong Kook},
  booktitle={DCASE 2018 workshop},
  year={2018}
}

@article{shim2020audio,
  title={Audio tagging and deep architectures for acoustic scene classification: Uos submission for the DCASE 2020 challenge},
  author={Shim, HJ and Kim, JH and Jung, JW and Yu, Ha-jin},
  journal={Proceedings of the DCASE2020 Challenge, Virtually},
  pages={2--4},
  year={2020}
}

@article{cai2018exploring,
  title={Exploring the encoding layer and loss function in end-to-end speaker and language recognition system},
  author={Cai, Weicheng and Chen, Jinkun and Li, Ming},
  journal={arXiv preprint arXiv:1804.05160},
  year={2018}
}

@article{chung2020defence,
  title={In defence of metric learning for speaker recognition},
  author={Chung, Joon Son and Huh, Jaesung and Mun, Seongkyu and Lee, Minjae and Heo, Hee Soo and Choe, Soyeon and Ham, Chiheon and Jung, Sunghwan and Lee, Bong-Jin and Han, Icksang},
  journal={arXiv preprint arXiv:2003.11982},
  year={2020}
}

@inproceedings{ester1996density,
  title={A density-based algorithm for discovering clusters in large spatial databases with noise.},
  author={Ester, Martin and Kriegel, Hans-Peter and Sander, J{\"o}rg and Xu, Xiaowei and others},
  booktitle={kdd},
  volume={96},
  number={34},
  pages={226--231},
  year={1996}
}

\end{sloppy}
\end{document}